\documentclass[10pt,twocolumn,showpacs,longbibliography,amsmath,amssymb,osajnl,floatfix,superscriptaddress]{revtex4-1}

\usepackage{graphicx}
\usepackage{dcolumn}
\usepackage{bm}
\usepackage{color}
\usepackage{txfonts}
\usepackage{microtype}

\begin{document}

\author{Feng Wan}	\affiliation{MOE Key Laboratory for Nonequilibrium Synthesis and Modulation of Condensed Matter, School of Science, Xi'an Jiaotong University, Xi'an 710049, China}
\author{Rashid Shaisultanov}
\affiliation{Max-Planck-Institut f\"{u}r Kernphysik, Saupfercheckweg 1,
	69117 Heidelberg, Germany}
\author{Yan-Fei Li}	\affiliation{MOE Key Laboratory for Nonequilibrium Synthesis and Modulation of Condensed Matter, School of Science, Xi'an Jiaotong University, Xi'an 710049, China}		
\author{Karen Z. Hatsagortsyan}\email{k.hatsagortsyan@mpi-hd.mpg.de}
\affiliation{Max-Planck-Institut f\"{u}r Kernphysik, Saupfercheckweg 1,
	69117 Heidelberg, Germany}
\author{Christoph H. Keitel}
\affiliation{Max-Planck-Institut f\"{u}r Kernphysik, Saupfercheckweg 1,
	69117 Heidelberg, Germany}
\author{Jian-Xing Li}\email{jianxing@xjtu.edu.cn}
\affiliation{MOE Key Laboratory for Nonequilibrium Synthesis and Modulation of Condensed Matter, School of Science, Xi'an Jiaotong University, Xi'an 710049, China}

\title{Ultrarelativistic polarized positron jets  via collision of  electron and  ultraintense laser beams }

\date{\today}

\begin{abstract}

Relativistic spin-polarized positron beams are indispensable for future electron-positron colliders to test  modern high-energy physics theory with high precision. However, present techniques require very large scale facilities for those experiments.
 We put forward a novel efficient method for generating  ultrarelativistic polarized positron beams employing currently available laser fields. For this purpose the generation of polarized positrons via multiphoton Breit-Wheeler pair production  and the associated spin dynamics in single-shot interaction of an ultraintense laser pulse with an ultrarelativistic electron beam is investigated in the quantum radiation-dominated regime. The pair production spin asymmetry in strong fields,  significantly exceeding the asymmetry of the radiative polarization, produces locally highly polarized particles, which are split by  a specifically tailored small ellipticity of the laser field into two oppositely polarized beams along the minor axis of laser polarization. In spite of  radiative  de-polarization, a dense positron beam with up to about 90\% polarization can be generated in tens of femtoseconds.
The method may eventually usher high-energy physics studies into smaller-scale laser laboratories.
 

\end{abstract}

\maketitle

\section{Introduction}

Relativistic polarized positron beams complemented with polarized electron beams are fundamental experimental tools to test symmetry properties in physics, in particular, in probing the structure of hadrons \cite{Maas_2004}, testing the Standard Model \cite{Androic_2013}, and searching for  new physics beyond the Standard Model \cite{Moortgat2008}. High-energy electrons and positrons can be directly polarized in a storage ring via radiative polarization (Sokolov-Ternov effect) \cite{Sokolov_1964,Sokolov_1968,Baier_1967,Baier_1972,Derbenev_1973}, which requires a rather long polarization time (typically from minutes to hours), because the magnetic fields of a synchrotron are too weak. In non storage ring facilities polarized positrons (electron-positron pairs) can be obtained   in a high $Z$-target by circularly polarized (CP) high-energy $\gamma$-photons \cite{Variola_2014}. However,  the latter have to be first produced from Compton backscattering of a CP laser light on a few-GeV electron beam \cite{Hirose_2000,Omori_2006}, synchrotron radiation of a multi-GeV electron beam travelling through a helical undulator \cite{Alexander_2008,McDonald_2009,Baynham_2011}, or the bremsstrahlung of polarized high-energy electrons \cite{Abbott_2016}. In these methods, however, the photon luminosity is low and requires a large
amount of repetitions or shots to yield a dense positron beam.

Recently, the advanced strong laser techniques, with intensities of the order of $10^{19}$-$10^{21}$ W/cm$^2$,  have been applied for generation of electron-positron jets in laser-solid interaction \cite{Chen_2009,Chen_2010,Chen_2013,Chen_2015a,Liang_2015,Chen_2016}, and electron-positron dense plasma jets \cite{Sarri_2015n} in laser-electron beam interaction. The electrons and positrons in those experiments are not polarized and aimed at modeling problems of laboratory astrophysics. In both setups initially produced $\gamma$-photons are converted into pairs via Bethe-Heitler process in a Coulomb field of high-$Z$ atoms.
Presently available petawatt-class lasers have capability for intensities  up to $10^{22}$ W/cm$^2$ \cite{Yanovsky2008,Vulcan}, and more are envisaged in near future \cite{Zou_2015,ELI,Exawatt}.  In such strong laser fields QED processes become nonlinear involving multiphoton processes \cite{Goldman_1964, Nikishov_1964, Brown_1964, Ritus_1985}, which, in particular, allow for electron-positron pair production due to direct interaction of a  $\gamma$-photon with a strong laser field (nonlinear Breit-Wheeler (BW) process) \cite{Ritus_1985}, and the  $\gamma$-photon generation is enhanced in the nonlinear Compton scattering regime. There are many proposals to generate unpolarized electron-positron beams in the nonlinear QED regime, see \cite{Hu_2010, Sarri_2013, Ridgers_2013,Ridgers_2014, Nakamura_2015,Blackburn_2017,
 Olugh_2019} and references therein, and even avalanche-like electromagnetic cascades in the case of future extreme laser intensities $ \gtrsim 10^{24}$~W/cm$^2$, see \cite{Bell_2008, Piazza2012, Marklund2006, Mourou2006} and references therein. Although the laser  magnetic field  can be much stronger (of the order of $10^5$ T) than the synchrotron magnetic field (of order 1 T), the radiative polarization  with laser fields is suppressed due to the symmetric character of the field \cite{Kotkin_2003,Ivanov_2004,Karlovets_2011,Seipt_2018}, i.e., the particles in adjacent half-cycles are polarized oppositely. Attractiveness of a strong laser fields for particle polarization has been recently demonstrated in the case of a model laser field in the form of a strong rotating electric field  \cite{Sorbo_2017,Sorbo_2018}. Although, the rotating electric field models anti-nodes of the electric field of a circularly polarized standing laser wave, the electron bunch in such a field can be trapped only in nodes of the electric field \cite{Bashinov_2017}, and only few electrons may reach anti-nodes.

Nevertheless,  recently we have shown a way to polarize an electron beam  with currently available realistic laser fields \cite{li2019prl}. Nonlinear interaction of electrons with an elliptically polarized (EP) laser field has been shown to result in the Stern-Gerlach type of splitting of the beam with respect to polarization due to the spin dependence of radiation reaction. The latter is a consequence of the asymmetry of the photon emission probabilities with respect to the electron spin in the given external fields. Furthermore, in strong external fields the electron-positron pair production probabilities possess much higher asymmetry with respect to the spin of the created particles than the radiation probabilities in the same field. The latter property is harnessed in this investigation for generation of highly polarized positrons.

\begin{figure}
	\setlength{\abovecaptionskip}{-0.0cm} 	
\includegraphics[width=1\linewidth]{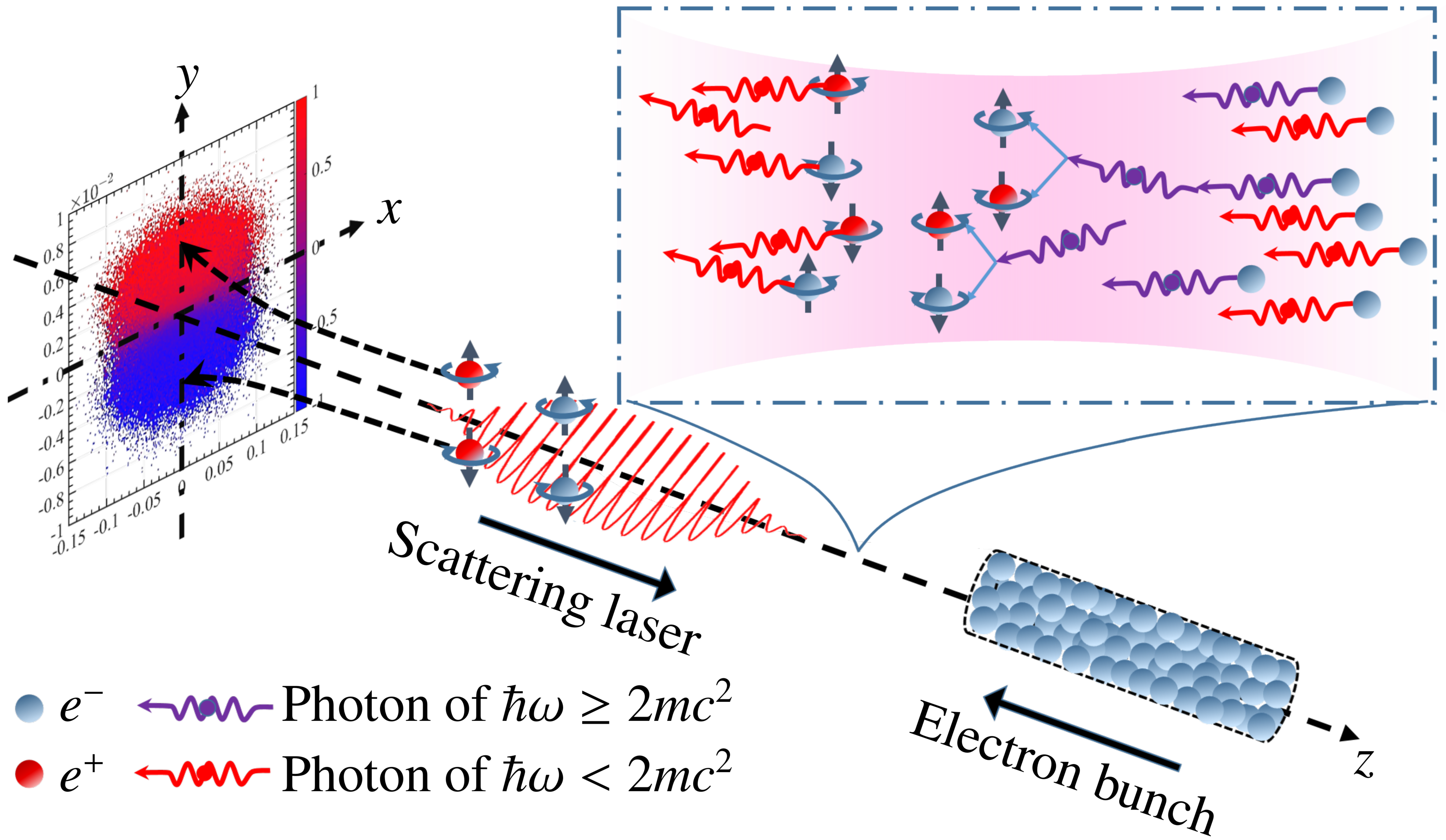}
	\caption{Scenario of generation and polarization of a positron beam. An ultraintense EP laser pulse propagates along $+z$ direction, with the major axis of the polarization ellipse along $x$ axis, and head-on collides with an ultrarelativistic electron bunch. 
	The radiated high-energy $\gamma$-photons further decay into polarized electron-positron pairs with respect to the instantaneous SQA, see inset. Further, the particles are split into two beams along $y$ axis with respect to the spin projection (red $S_y>0$, blue $S_y<0$), because of  asymmetries of spin-dependent pair-production and photon-emission probabilities.}
	\label{fig1}
\end{figure}

In this work, generation and  polarization of a positron (electron-positron) beam in the interaction of an ultraintense EP laser pulse with a counterpropagating ultrarelativistic electron bunch have been investigated in the quantum radiation-dominated regime, see the interaction scenario in Fig.~\ref{fig1}. The laser intensity is strong enough, such that $\gamma$-photons generated due to nonlinear Compton scattering can produce electron-positron pairs via multiphoton BW process during further interaction with the laser pulse. Due to large spin asymmetry of the pair production process, the produced positrons are highly polarized along the instantaneous magnetic field in their rest frames, which we choose as the spin quantization axis (SQA).
With a proper choice of ellipticity of the laser field, the positrons are split  into two beams along $y$ axis with respect to the spin projection, because of asymmetries of spin-dependent
pair-production and photon-emission probabilities.
The pair production asymmetry is dominating, which brings about much larger polarization of positrons in separated beams, compared with the radiative polarization of incoming electrons known from \cite{li2019prl}.
We underline that in our scheme the laser field is not asymmetric, and asymmetry of the pair production probability is reflected in the angular separation of the oppositely polarized parts of the beam. This is in contrast to \cite{Chen_2019}, where asymmetric two-color laser field is applied for positron polarization, yielding though considerable less polarization and larger angular spreading. A similar two-color laser model is proposed to polarize electrons as well \cite{Song_2019, Seipt_2019}.
Our theoretical analysis is based on  Monte Carlo simulations of the particles' spin and space-time dynamics taking into account the radiation and pair production processes in spin resolved manner.

The laser and electron beam parameters are chosen such as to have significant pair production with $\chi_e\sim\chi_\gamma\sim 4$, and significant radiation reaction in the radiation-dominated regime with $R\equiv \alpha\xi\chi_e\geq 1$ (the electron radiation losses during a laser period are comparable with or larger than the electron initial energy \cite{Koga2005,Piazza2012}). Here, we define
the
invariant laser field parameter $\xi\equiv |e|E_0/(m\omega_0 c)$, the nonlinear QED parameters  $\chi_e\equiv |e|\hbar\sqrt{-(F_{\mu\nu}p_e^{\nu})^2}/m^3c^4$ (for  electrons) and $\chi_{\gamma}\equiv |e|\hbar\sqrt{-(F_{\mu\nu}k_{\gamma}^{\nu})^2}/m^3c^4$ (for $\gamma$-photons) \cite{Ritus_1985},
$\alpha$ the fine structure constant,  $c$  the speed of the light in vacuum, $\hbar$  the  Planck constant, while $F_{\mu\nu}$, $E_0$ and $\omega_0$ are the field tensor, the amplitude and the frequency of the laser field, respectively, $e$ and $m$ charge and mass of the electron, respectively,   and,  $p_e=(\varepsilon_e/c,{\bf p}_e)$ and $k_\gamma$ the 4-momenta of the  electron and the $\gamma$- photon, respectively.  When the electron counterpropagates with the laser beam,  $\chi_e\approx 2(\hbar\omega_0/mc^2)\xi\gamma_e$, where $\gamma_e$ is the electron's Lorentz factor.  \\

\section{Simulation method}

\textbf{The Monte-Carlo method for  pair polarization.} We develop a Monte-Carlo method to model spin effects during electron-positron pair production and propagation in arbitrary electromagnetic fields by employing spin-resolved probabilities of radiation and pair production in the local constant field approximation (LCFA), valid at $\xi \gg 1$ \cite{Ritus_1985, Baier1998}.
 We employ the spin-resolved probabilities in LCFA for photon emission, see Eq.~(1) in \cite{li2019prl}, and for pair production the following formula, derived in the leading order contribution with respect to $1/\gamma_e$ through the QED operator method of Baier-Katkov \cite{Baier_1973}:

\begin{widetext}
\begin{eqnarray}\label{pairspin}
\frac{{\rm d^2}W_{pair}}{{\rm d}\varepsilon_{+}{\rm d}t}&=&W_P\frac{\varepsilon_{-}}{\varepsilon_{+}}\left\{\left(1-\frac{\varepsilon_{+}}{\varepsilon_{-}}\right)^2\left[2{\rm K}_{\frac{2}{3}}(\rho)-{\rm IntK}_{\frac{1}{3}}(\rho)\right](1+{\bf S}_{+}\cdot{\bf S}_{-})+\frac{\varepsilon_{\gamma}^2}{\varepsilon_{-}^2}\left[2{\rm K}_{\frac{2}{3}}(\rho)+{\rm IntK}_{\frac{1}{3}}(\rho)\right](1-{\bf S}_{+}\cdot{\bf S}_{-})+2\left(1-\frac{\varepsilon_{+}^2}{\varepsilon_{-}^2}\right)({\bf S}_{+}+{\bf S}_{-})\right.\nonumber\\
&&\left.\left[{\bm\beta}_{-}\times\hat{{\bf a}}_{-}\right]{\rm K}_{\frac{1}{3}}(\rho)-2\frac{\varepsilon_{\gamma}^2}{\varepsilon_{-}^2}({\bf S}_{-}-{\bf S}_{+})\left[{\bm\beta}_{-}\times\hat{\bf a}_{-}\right]{\rm K}_{\frac{1}{3}}(\rho)+4\left[\left(\frac{\varepsilon_+^2}{\varepsilon_-^2}+1\right)\rm {IntK}_{\frac{1}{3}}(\rho)-\left(\frac{\varepsilon_+}{\varepsilon_-}-1\right)^2\rm {K}_{\frac{2}{3}}(\rho)\right]({\bf S}_{+}\cdot\bm\beta_{-})({\bf S}_{-}\cdot\bm\beta_{-})+\right.\nonumber\\
&&\left. 2\frac{\varepsilon_{\gamma}^2}{\varepsilon_{-}^2}({\bf S}_{+}\cdot{\bf S}_{-}){\rm IntK}_{\frac{1}{3}}(\rho) \right\},
\end{eqnarray}
\end{widetext}
where,$W_P\equiv\alpha m^2 c^4/(16\sqrt{3}\pi\hbar\varepsilon_{\gamma}^2)$, $\varepsilon_{\gamma}$, $\varepsilon_{-}$ and $\varepsilon_{+}$ are energies of the $\gamma$- photon, electron and positron, respectively, with $\varepsilon_{\gamma}=\varepsilon_{-}+\varepsilon_{+}$, and $\rho=2\varepsilon_{\gamma}^2/(3\chi_{\gamma}\varepsilon_{+}\varepsilon_{-})$, ${\bf S}_{-}$ and ${\bf S}_{+}$ are spin vectors of electron and positron, respectively, ${\bm \beta}_{-}$ is the electron velocity scaled by $c$,
$\hat{\bf a}_{-}={\bf a}_{-}/|{\bf a}_{-}|$ with ${\bf a}_{-}$ the electron acceleration, ${\rm IntK}_{\frac{1}{3}}(\rho)\equiv\int_{\rho}^{\infty} {\rm d}x {\rm K}_{\frac{1}{3}}(x)$, and ${\rm K}_n$ is the $n$-order modified Bessel function of the second kind.
Note that the probability in Eq.~(\ref{pairspin}) is summed up by photon polarization.
Summing over ${\bf S}_{+}$ and ${\bf S}_{-}$ in Eq.~(\ref{pairspin}), the widely employed spin averaged pair production probability  \cite{Bell_2008, Piazza2012} is obtained:
\begin{eqnarray}\label{pair}
\frac{{\rm d^2}\overline{W}_{pair}}{{\rm d}\varepsilon_{+}{\rm d}t}&=&16W_P\left\{{\rm IntK}_{\frac{1}{3}}(\rho)+\frac{\varepsilon_{+}^2+\varepsilon_{-}^2}{\varepsilon_{+}\varepsilon_{-}}{\rm K}_{\frac{2}{3}}(\rho)\right\}.
\end{eqnarray}
 If summing over only ${\bf S}_{+}$ or ${\bf S}_{-}$ in Eq.~(\ref{pairspin}), the pair production probability solely depending on ${\bf S}_{-}$ or ${\bf S}_{+}$ is obtained:
\begin{eqnarray}
\label{espin}
\frac{{\rm d^2}W_{pair}^\mp}{{\rm d}\varepsilon_{+}{\rm d}t}=\frac{1}{2}\frac{{\rm d^2}\overline{W}_{pair}}{{\rm d}\varepsilon_{+}{\rm d}t} \mp 8W_P\frac{\varepsilon_{\gamma}}{\varepsilon_{\mp}}\left[{\bm\beta}_{-}\times\hat{{\bf a}}_{-}\right]{\bf S}_{\mp}{\rm K}_{\frac{1}{3}}(\rho).
\end{eqnarray}

The stochastic pair polarization effects are carried out by the following procedure, following the spirit of the quantum jump approach \cite{Molmer_1996,Plenio_1998}.
Three random numbers, $N_r$, $N'_r$ and $N''_r$  in $[0, 1]$, are used. First, at each pair formation length, as the spin-free pair-production probability in Eq.~(\ref{pair}) $\overline{W}_{pair}\geq N_r$, a pair is produced. Then, one of ${\bf S}_{-}$ and ${\bf S}_{+}$, e.g., ${\bf S}_{-}$, is first determined: ${\bf S}_{-}$ is stochastically collapsed into one of its basis states defined with respect to the SQA, which is chosen
along the magnetic field in the rest frame of the electron (along ${\bm\beta}_{-}\times\hat{{\bf a}}_{-}$). In particular, ${\bf S}_{-}$ is either parallel (spin-up) or anti-parallel (spin-down) to its instantaneous SQA with probabilities $W_{pair}^{-\uparrow}$ and $W_{pair}^{-\downarrow}$, respectively. Here, $\overline{W}_{pair}=W_{pair}^{-\uparrow}+W_{pair}^{-\downarrow}$, and $W_{pair}^{-\uparrow}$ and $W_{pair}^{-\downarrow}$ are calculated via Eq.~(\ref{espin}).  If $W_{pair}^{-\uparrow}/{W}_{pair}\geq N'_r$, ${\bf S}_{-}$ is up, otherwise, down. Finally, since ${\bf S}_{-}$ and $W_{spin}^-$ in Eq.~(\ref{espin}) are already known, $W_{spin}$ in Eq.~(\ref{pairspin}) becomes solely dependent on ${\bf S}_{+}$, which is also either parallel (spin-up) or anti-parallel (spin-down) to its instantaneous SQA (anti-parallel to that of the electron)  with probabilities  $W_{pair}^{\uparrow}$ and $W_{pair}^{\downarrow}$, respectively. Here, $W_{pair}^-=W_{pair}^{\uparrow}+W_{pair}^{\downarrow}$.  If $W_{pair}^{\uparrow}/{W}_{pair}^-\geq N''_r$, ${\bf S}_{+}$ is set  up, otherwise, down.

Since the electron or positron propagates in the  external laser field, after a photon emission, the spin state is assumed to stochastically collapse into  its instantaneous SQA employing the spin-resolved probabilities of photon emission \cite{li2019prl}.
Between photon emissions, its dynamics is described by Newton equations, and the spin precession is governed by the Thomas-Bargmann-Michel-Telegdi  equation
\cite{Thomas_1927,Bargmann_1959,supplemental}. \\

\begin{figure}[t]
	
		\setlength{\abovecaptionskip}{-0cm}
	\includegraphics[width=1.0\linewidth]{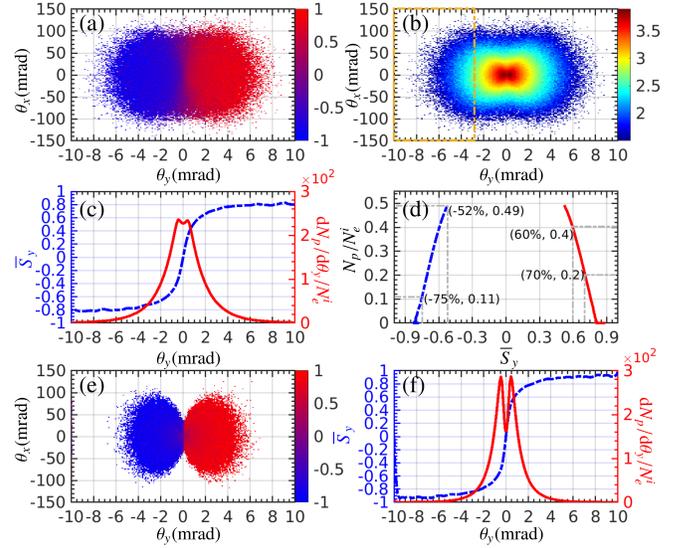}
	\caption{
		(a) and (b): Transverse distributions of the positron spin component $S_y$ and normalized density $\log_{10}\left( {\rm d}^2 N_p/{\rm d}\theta_x{\rm d}\theta_y/N_e^i \right) {\rm rad}^{-2}$ with respect to the deflection angles $\theta_x=\arctan(p_{+, x}/p_{+, z})$ and $\theta_y=\arctan(p_{+, y}/p_{+, z})$.
		(c): Average spin $\overline{S}_y$ (blue-dashed curve) and positron density $\textrm{d}N_p/\textrm{d}\theta_y/ N_e^i$ (red-solid curve) vs $\theta_y$; $N_p$ and $N_e^i$ are the number of positrons and primary electrons, respectively. (d): $N_p/N_e^i$ vs $\overline{S}_y$. The red-solid and blue-dashed curves represent the positron polarization parallel and anti-parallel to the $+y$ axis, respectively.  Positrons corresponding to the coordinate (-75\%, 0.11) are indicated in the yellow-dashed box in (b). (e) and (f) show the same information as  (a) and (c), respectively, but artificially exclude radiation effects of the positrons. The laser and electron beam parameters are given in the text.
	}
	\label{fig2}
\end{figure}

\section{Results and analysis}

\textbf{Positron polarization.} Polarization effects of created positron beam are illustrated in Fig.~\ref{fig2}, and those of the electron beam in \cite{supplemental}. We employ a realistic tightly-focused EP laser pulse with a Gaussian temporal profile, and the spatial distribution of the electromagnetic fields takes into account up to $\epsilon_0^3$-order of the nonparaxial solution, where $\epsilon_0=w_0/z_r$, while $w_0$ is the laser focal radius, $z_r=k_0w_0^2/2$ the Rayleigh length with laser wave vector $k_0=2\pi/\lambda_0$, and $\lambda_0$ the laser wavelength \cite{Salamin2002, supplemental}. The laser peak intensity  $I_0\approx1.37\times10^{22}$ W/cm$^2$ ($\xi=100$), wavelength $\lambda_0=1$~$\mu$m, pulse duration $\tau = 8T_0$ with the period $T_0$, focal radius $w_0=5$ $\mu$m, and ellipticity $\epsilon=|E_y|/|E_x|=0.03$. A  cylindrical electron bunch is considered, with the radius  $w_e= \lambda_0$, length $L_e = 5\lambda_0$, and density $n_e^i\approx 6.4\times10^{16}$ cm$^{-3}$. The transverse electron distribution is Gaussian and the longitudinal one is uniform. The collision polar angle with respect to the laser propagation direction is $\theta_e=180^{\circ}$ and the azimuthal angle $\phi_e=0^{\circ}$. The angular divergence of the electron beam is 0.3 mrad, the initial kinetic energy $\varepsilon_0=10$ GeV, and the energy spread $\Delta \varepsilon_0/\varepsilon_0 =0.06$. The electron beam with such parameters  can be obtained by multistage coupling of independent laser-plasma accelerators \cite{Steinke_2016, Cros_2016} or laser wakefield accelerators \cite{Leemans2014,Leemans_2019}. The pair production and radiation reaction are significant at these parameters as
$\chi_e^{\rm max}\approx4.9$ and $\chi_\gamma^{\rm max}\approx 4.6$, and $R\approx 4$, but avalanche-like electromagnetic cascades are suppressed.

The positrons are polarized and split by propagation direction into two beams polarized parallel and anti-parallel to  the $+y$ axis, respectively, with a splitting angle of about 10 mrad, see Fig.~\ref{fig2}(a). The splitting angle is much larger than the beam angular divergence ($\sim1/\gamma_{+}<1$ mrad, with the positron Lorentz gamma-factor $ \gamma_{+}$) \cite{supplemental}, and the angular resolution of the current technique for electron detectors (less than 0.1 mrad) \cite{Wang2013,Leemans2014,Wolter2016,Chatelain2014}.
 The positrons mainly concentrate around the beam center, since  the transverse ponderomotive force is relatively small, see Fig.~\ref{fig2}(b), and the slight split in positron density corresponds to the split of the parent  $\gamma$-photons, which are emitted during electron spin-dependent dynamics \cite{supplemental}.

As shown in Fig.~\ref{fig2}(c), near $\theta_y=0$, the positron density is rather high, but the average spin $\overline{S}_y$ is relatively low. With the increase of $|\theta_y|$, the positron density declines, however, $\overline{S}_y$ remarkably ascends until about 80\%.

To obtain a polarized positron beam, one has to implement a selection over $\theta_y$, i.e., choose $0<\theta_{y0}< \theta_y$ to select the spin-up polarization (or $0>\theta_{y0}> \theta_y$, for the spin-down polarization). Figure~\ref{fig2}~(d) shows the relative number of positrons vs the average spin over the beam, at varying the value of $\theta_{y0}$.  When splitting the beams at $\theta_{y0}=0$, one obtains polarized beams of $|\overline{S}_y|\approx 52\%$, with $N_p/N_e^i=0.49$, see the coordinate (-52\%, 0.49) in Fig.~\ref{fig2}(d). Since the polarization of primary electron beam  due to  radiative spin effects  is parallel to that of the created electrons  and relatively low, the total polarization of electron beam  is about 10\% lower than that of the positron beam \cite{supplemental}. Moreover, the polarization  dramatically increases as the positrons near $\theta_y=0$ are excluded (with increasing $|\theta_{y0}|$), e.g., 11\% positrons in the $\theta_y$ region indicated by the dashed box in Fig.~\ref{fig2}(b), corresponding to a splitting angle $\theta_{y0}\approx -3$~mrad, have an average polarization of about -75\%.

We underline that the positrons are mainly polarized due to the multiphoton BW process, and the polarization is depressed by the stochastic radiative spin effects.
As revealed in Figs.~\ref{fig2}(e) and (f), when radiation effects are artificially removed, the positrons are more polarized and concentrated.

We have analyzed the robustness of the polarization scheme, via considering the cases of larger energy spread $\Delta \varepsilon_0/\varepsilon_0=0.1$, larger angular divergence of 1~mrad, and different collision angles $\theta_e=179^\circ$ and $\phi_e=90^\circ$. In all cases stable and uniform results are obtained \cite{supplemental}.

\begin{figure}[t]
		\setlength{\abovecaptionskip}{-0cm}
	\includegraphics[width=1\linewidth]{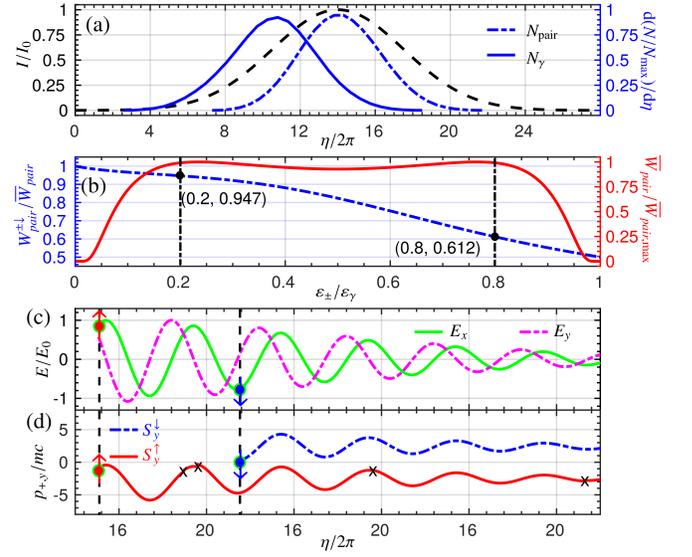}
	\caption{(a): Normalized parent $\gamma$-photon emission
rate (blue-solid), pair production
rate (blue-dash-dotted) and laser pulse intensity (black-dashed) vs the laser phase $\eta=\omega_0 t- k_0 z$.
		(b): Normalized pair production probability  (red-solid) and probability of positron or electron polarizing anti-parallel to its SQA (blue-dash-dotted) vs $\varepsilon_{\pm}/\varepsilon_\gamma$.
	 (c) Normalized field components $E_x$ (green-solid) and $E_y$ (magenta-dashed); (d): Positron momentum, the red-up and blue-down arrows indicate the spin being parallel and anti-parallel to  $+y$ axis,respectively. The color points indicate pair creation, and the black crosses  photon emission.
	}
	\label{fig3}
\end{figure}

\textbf{Physical interpretation.} The reasons for the positron beam polarization and splitting are analyzed in Fig.~\ref{fig3}.  Parent $\gamma$- photons are mainly emitted  at the front part of the laser pulse, and subsequently produce pairs near the laser pulse peak since there $\chi_\gamma\sim\xi$ is large. Average phase delay between parent photon emission and pair production is about 3.5 periods, coinciding with the estimation of the mean free path  $\overline{\lambda}\approx\lambdabar_c\varepsilon_{\gamma}/[0.16\alpha m c^2 {\rm K}_{1/2}^2(2/3\chi_{\gamma})]$ \cite{Erber_1966}, see Fig.~\ref{fig3}(a).

The pair production probability $\overline{W}_{pair}$ normalized to its maximum value and the probability $W_{pair}^{\pm \downarrow}$ of the polarized positron (electron) creation,
anti-parallel to its SQA, are shown in Fig.~\ref{fig3}(b). The pair production probability (red-solid curve) is the largest within the energy interval $0.2\lesssim \varepsilon_{\pm}/\varepsilon_\gamma\lesssim 0.8$, with the positron  $\varepsilon_{\pm}$ and $\gamma$-photon $\varepsilon_\gamma$ energies, where  the spin of the positron (electron) is anti-parallel to its SQA with a high probability $0.612\lesssim W_{pair}^{\pm \downarrow}/\overline{W}_{pair}\lesssim 0.947$ (blue-dashed curve). Thus, the pairs are created with a preferable anti-parallel polarization with respect to the SQA.

The positron beam splitting is analyzed in Fig.~\ref{fig3}(c) and (d).
In the considered left-handed EP laser pulse, with the major axis of elliptical polarization along $x$-axis, the electric field $E_x$-component (magenta-dashed) has a $\pi/2$ phase delay with respect to  $E_y$ (green-solid), and
the vector potential ${\bf A}(\eta)$  is delayed by $\pi/2$ with respect to the field ${\bf E}(\eta)$. Thus,
$A_{y}$ and $E_x$ are oscillating in opposite phase.
In the laser field the SQA is along ${\bm\beta}_{\pm}\times\hat{{\bf a}}_{\pm}\propto \pm{\bm\beta}_{\pm}\times{\bf E}\pm{\bm\beta}_{\pm}\times({\bm \beta}_{\pm}\times{\bf B})\sim \pm(1-\beta_{\pm, z}){\bm\beta}_{\pm}\times{\bf E}$ \cite{supplemental}, with the laser electric ${\bf E}$ and magnetic ${\bf B}$ fields, the positron (electron) velocities $\textbf{v}_\pm$ and accelerations $\textbf{a}_\pm$, and ${\bm\beta}_{\pm}=\textbf{v}_\pm/c$. As ${\bm\beta}_{\pm, z}$ is along $-z$ direction, the SQA sign of the positron (electron) is opposite to (the same as) the sign of $E_x$. In the employed EP laser field, the pair production mostly occurs at maxima
of $|E_x|$.

If a pair is created at $E_x>0$  with $\eta_+$ (red point in Fig.~\ref{fig3}(c)), the SQA of the positron is in $-y$ direction, and the positron spin is very probably in $+y$ direction (anti-parallel to its SQA) indicated by red-up arrow. The corresponding momentum $p_{+, y}=p_{+, y}^i +eA_y(\eta_+)-eA_y(\eta)$, with the primary momentum of positron $p_{+, y}^i\sim -e A_y(\eta_\gamma)$ inherited from its parent $\gamma$-photon created at $\eta_\gamma$,  is shown in Fig.~\ref{fig3}(d). The positron final momentum is $p_{+, y}^f=p_{+, y}^i+eA_y(\eta_+)$. As the positron is created at the  peak of $E_x(\eta_+)>0$, the $y$-component of the vector potential is at the negative peak $A_y(\eta_+)<0$. Moreover, $p_{+, y}^i\ll eA_y(\eta_+)$, because the $\gamma$-photons are created at much lower laser intensities than the pairs, and correspondingly, $|A_y(\eta_\gamma)|\ll |A_y(\eta_+)|$, see Fig.~\ref{fig3}(a)). Therefore, $p_{+, y}^f\approx eA_y(\eta_+)<0$.
Thus, the spin-up positron moves in $-y$ direction, and $\theta_y=\arctan(p_{+,y}/p_{+,z})>0$ (as known $p_{+,z}<0$). The similar analysis applies for the positron created at
$E_x<0$: since the spin-down positron indicated in blue-down arrow first experiences acceleration by $E_y$, it finally moves to $+y$ direction, and relevant $\theta_y<0$. See Fig.~\ref{fig2}(a).
In linearly  and circularly polarized laser fields the discussed angular splitting of polarized positrons  cannot take place \cite{supplemental}.

\begin{figure}[t]
		\setlength{\abovecaptionskip}{-0cm}
	\includegraphics[width=1.0\linewidth]{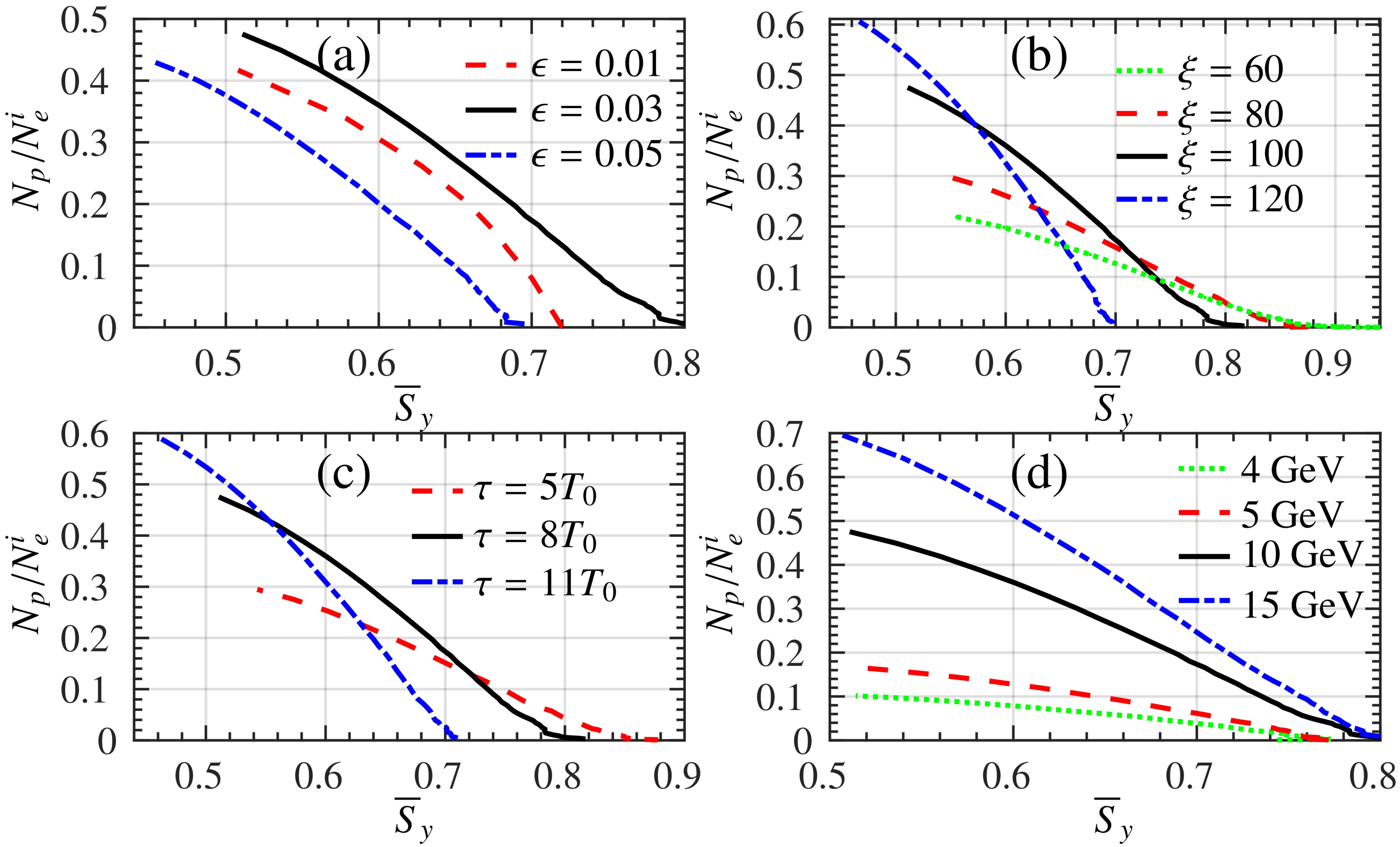}
	\caption{ (a)-(d):
		Impacts of ellipticity, intensity and pulse duration of the laser, and mean energy of the primary electrons  $\varepsilon_0$ on the polarization, respectively. Other parameters are the same as those in Fig.~\ref{fig2}.
	}
	\label{fig4}
\end{figure}

After creation the positrons still move in the strong laser field and  can emit photons. The photon emissions
induce momentum spreading and stochastic spin flips \cite{li2019prl}, and consequently, depress the polarization, cf. Fig.~\ref{fig2} (a),(c) with (e),(f), which necessitate to restrict the laser pulse duration,
see also Fig.~\ref{fig4}(c). In particular, in a monochromatic laser wave
the positron beam cannot be polarized because of essential polarization damping induced by photon emissions
\cite{supplemental}.

\textbf{Impact of the laser and electron parameters on the positron polarization.} The results of investigations of the impact of the laser and electron beam parameters are presented in Fig.~\ref{fig4}.
 First, the ellipticity $\epsilon$ is a very crucial parameter. If $\epsilon$ is too small, the splitting angle $\theta_s\sim p_{y}/p_{x}\propto E_y/E_x$ is very small as well, and the polarized positrons partially overlap near $p_y=0$, which reduces the degree of polarization (cf. the ultimate case of linear polarization).
 Oppositely,  largely increasing ellipticity can increase the splitting angle, but unfortunately also the SQA rotation (cf., the ultimate case of circular polarization). As a result the average polarization decreases, see Fig.~\ref{fig4}(a). The  optimal  ellipticity is of order of $10^{-2}$ to $10^{-1}$. The trade off exists also for the laser intensity, pulse duration, and the electron energy. From one side, the considered effect relies on pair production and requires large $\chi_\gamma\sim\xi\varepsilon_{\gamma}/mc^2\gg 1$ and much pair creation. From another side, the stochastic radiative spin flips during the positron propagating through the laser field   smear out the considered effect which imposes restriction on the photon emissions.
 For this reason,
 with increasing $\xi$ and $\tau$, the positron number, $N_p\propto N_{\gamma}\sim\alpha\xi\tau/T_0$, is enhanced, but the polarization is depressed, see Figs.~\ref{fig4}(b) and (c). While increasing the primary electron mean energy $\varepsilon_0$, the positron number is enhanced as $\chi_\gamma$ increases, however, the polarization is not influenced significantly, see Fig.~\ref{fig4}(d). When proper laser and electron beam parameters are employed, the high polarization up to 90\% can be achieved, e.g., as shown in green-dotted curve in Fig.~\ref{fig4}(b), about 1.14\% positrons can reach a polarization of about 85.76\%.\\

\section{Conclusion}

We have developed a Monte-Carlo method for simulating positron polarization via electron-positron pair production process in strong laser fields. Our investigation shows that by adding a small ellipticity to the strong laser field, it is possible to achieve angular splitting of the created positrons with respect to the polarization, and in this way to obtain highly polarized dense positron beams. In particular, with  currently available laser technique it is possible to achieve about 86\% (even up to 90\%) polarization of the positron beam, with the number of positrons more than 1\% of the initial electrons. Generally, larger polarization can be obtained at the expense of decreasing the number of positrons in the beam. The considered polarization effect is shown to be robust with respect to the laser and electron beam parameters. The radiation accompanying pair production induces spin flips and because of that reduces the positron polarization. To avoid the negative role of photon emissions, one should trade-off the laser pulse duration and intensity.  The optimal parameters include a laser intensity of the order of $10^{22}$~W/cm$^2$, an ellipticity of the order of $10^{-2}$ to $10^{-1}$,  a laser pulse duration less than about 10 cycles, and an initial electron energy of several GeVs.
Combining the proposed method with the laser-wakefield electron acceleration technique will allow an all-optical way for generating polarized ultrarelativistic positron beams, and the polarization of such laser-driven electron beam can be measured via a polarimetry method of nonlinear Compton scattering \cite{Li_2019spin}.\\

 \textbf{Acknowledgements}\\
 
We are grateful to A. Di Piazza, M. Tamburini, and Y.-Y. Chen for helpful discussions. This work is supported by the National Natural Science Foundation of China (Grants Nos. 11874295, 11804269), the  Science Challenge Project of China (No. TZ2016099), and the National Key Research and Development Program of China (Grant No. 2018YFA0404801). \\

 \textbf{References}\\
 
\bibliography{QEDspin}

\end{document}